\documentclass[aps,a4paper,10pt,floatfix]{revtex4-2}
\usepackage{amsmath,amssymb}

\usepackage[dvipdfmx]{graphicx}
\usepackage{color}

\newcommand{\red}[1]{}

\newcommand{\tphi}{\tilde{\phi}}

\newcommand{\be}{\begin{equation}}
\newcommand{\ee}{\end{equation}}
\newcommand{\bi}{\begin{itemize}}
\newcommand{\ei}{\end{itemize}}

\begin{document}

\title{Observability transitions in clustered networks}
\date{\today}

\author{Takehisa Hasegawa}
\email{takehisa.hasegawa.sci@vc.ibaraki.ac.jp}
\affiliation{Graduate School of Science and Engineering, Ibaraki University, 2-1-1, Bunkyo, Mito, Japan 310-8512}
\author{Yuta Iwase}
\affiliation{Graduate School of Science and Engineering, Ibaraki University, 2-1-1, Bunkyo, Mito, Japan 310-8512}

%%%%%%%%%%%%%%%%%%%%%%%%%%%%%%%%%%%%%%%%%%%%%%%%%%%%%%%%%%%%%%%%%%%%%%%%%%%
\begin{abstract}
We investigate the effect of clustering on network observability transitions. 
In the observability model introduced by Yang, Wang, and Motter [Phys. Rev. Lett. 109, 258701 (2012)], a given fraction of nodes are chosen randomly, and they and those neighbors are considered to be observable, while the other nodes are unobservable. 
For the observability model on random clustered networks, we derive the normalized sizes of the largest observable component (LOC) and largest unobservable component (LUC).
Considering the case where the numbers of edges and triangles of each node are given by the Poisson distribution, we find that both LOC and LUC are affected by the network's clustering: more highly-clustered networks have lower critical node fractions for forming macroscopic LOC and LUC, but this effect is small, becoming almost negligible unless the average degree is small. We also evaluate bounds for these critical points to confirm clustering's weak or negligible effect on the network observability transition. The accuracy of our analytical treatment is confirmed by Monte Carlo simulations.
\end{abstract}
%%%%%%%%%%%%%%%%%%%%%%%%%%%%%%%%%%%%%%%%%%%%%%%%%%%%%%%%%%%%%%%%%%%%%%%%%%%

\maketitle

%%%%%%%%%%%%%%%%%%%%%%%%%%%%%%%%%%%%%%%%%%%%%%%%%%%%%%%%%%%%%%%%%%%%%%%%%%%
\section{Introduction}
%%%%%%%%%%%%%%%%%%%%%%%%%%%%%%%%%%%%%%%%%%%%%%%%%%%%%%%%%%%%%%%%%%%%%%%%%%%

{\it Dynamics on complex networks} is one of the main topics in network science \cite{barrat2008dynamical,newman2010networks,barabasi2016network,dorogovtsev2008critical,castellano2009statistical,pastor2015epidemic}. Over the last 20 years, many empirical studies have discovered that real-world networks often have common properties, such as the small-world property~\cite{watts1998collective} (where the mean shortest path length is proportional to the logarithm of the number of nodes and the clustering coefficient is relatively high) and the scale-free property~\cite{barabasi1999emergence} (where the degree distribution follows a power law). An enormous number of studies have already been devoted to investigating how such complex connections affect dynamics on networks.

The degree of clustering in a network is measured by its clustering coefficient $C$, which is the mean probability that two nodes connected to a common node are themselves connected. Many real-world networks have high clustering coefficients, and clustering is known to be a factor in various phenomenological models placed on networks, such as percolation~\cite{serrano2006percolation,newman2009random,miller2009percolation,gleeson2009bond,gleeson2010clustering,colomer2014double,hasegawa2020structure,mann2021percolation}, the spread of epidemics~\cite{newman2003properties,serrano2006percolation,britton2008epidemics,ball2010analysis,green2010large,mann2020two,mann2020cooperative}, information cascades~\cite{ikeda2010cascade,hackett2011cascades}, and synchronization~\cite{mcgraw2005clustering,xiang2006synchronizability}. However, the effect of clustering on the network observability model~\cite{yang2012network} is as yet unclear.

In~\cite{yang2012network}, Yang {\it et al.} introduced the network observability model as a simple model for power-grid systems controlled by phase measurement units, which we call sensors. In this model, the state of a given node is observable if there is a sensor on either that node or one of its neighbors, and is unobservable otherwise. In realistic systems, we want a small number of sensors to efficiently observe the nodes of a given network. Yang {\it et al.} considered this problem within an observability transition framework. They derived the size of the largest observable component (LOC), namely the largest connected component of observable nodes, for uncorrelated random networks with arbitrary degree distributions to determine the critical fraction of sensors required for the macroscopic LOC, and studied how this was affected by the network topology. 

The observability model has since been studied in various settings. Following~\cite{yang2012network}, the author (T.H.) and colleagues~\cite{hasegawa2013observability} studied it both analytically and numerically on correlated networks, showing that both uncorrelated networks and networks with negative degree correlations yielded larger LOCs than networks with positive degree correlations. Allard {\it et al.}~\cite{allard2014coexistence} treated the observability model on uncorrelated networks as a generalized percolation problem, showing the coexistence of macroscopic LOC and macroscopic largest unobservable component (LUC), which is the largest connected component of unobservable nodes. Other studies have examined observability transitions in networks with high betweenness preferences~\cite{shunkun2016observability} and the observability model in multiplex networks~\cite{osat2018observability}.

Recently, Yang and Radicchi~\cite{yang2016observability} extended the message-passing approach used for ordinary percolation transitions to describe LOC size in the observability model. Using nearly 100 real-world networks, they compared the message-passing approach's theoretical predictions with numerical simulation results, finding that this approach, which is valid as long as the network is locally treelike, produced almost perfect predictions in most cases, even for networks with very large clustering coefficients. This suggests that clustering has little effect on the observability transition.

Motivated by~\cite{yang2016observability}, we examine the effect of clustering on network observability in detail using the random clustered network model. Newman~\cite{newman2009random} and Miller~\cite{miller2009percolation} independently introduced a random graph model with clustering, many of whose network properties can be well described via generating function analysis. We use generating functions to calculate the LOC size for such networks, finding that it behaves differently depending on the clustering coefficient value, although this effect is small in most cases. We also obtain the LUC size, showing that the critical node fraction for the LUC also depends on the clustering coefficient, although this effect again becomes almost negligible when the average degree is large. In addition, we evaluate bounds on the critical node fractions for the LOC and LUC to confirm that clustering has a weak or negligible effect on the network observability transition. We also show that clustering has a negligible effect on network observability in scale-free networks, which supports the finding in~\cite{yang2016observability}. The results of our Monte Carlo simulations are in perfect agreement with these findings.

%%%%%%%%%%%%%%%%%%%%%%%%%%%%%%%%%%%%%%%%%%%%%%%%%%%%%%%%%%%%%%%%%%%%%%%%%%%
\section{Model}
%%%%%%%%%%%%%%%%%%%%%%%%%%%%%%%%%%%%%%%%%%%%%%%%%%%%%%%%%%%%%%%%%%%%%%%%%%%

In this study, we investigate the observability model using random clustered networks. The observability model, introduced by Yang {\it et al.}~\cite{yang2012network}, is defined as follows. For a given network with $N$ nodes, we place a sensor on each node with probability $\phi$. Nodes with sensors are directly observable, and all other nodes adjacent to at least one directly-observable node are indirectly observable. The remaining nodes, which are neither directly nor indirectly observable, are unobservable. This means all nodes are either directly observable (D), indirectly observable (I), or unobservable (U). 

Following previous studies~\cite{yang2012network,hasegawa2013observability,allard2014coexistence,shunkun2016observability,osat2018observability,yang2016observability}, we focus on the LOC, which is defined as the largest connected component consisting of D and I nodes. Similarly to the ordinary percolation, the LOC undergoes a phase transition at $\phi=\phi_c^{\rm LOC}$: it is small for $\phi<\phi_c^{\rm LOC}$, but becomes macroscopic for $\phi>\phi_c^{\rm LOC}$. Using the normalized LOC size $S_{\rm LOC}$, defined as the ratio of the LOC and network sizes, $S_{\rm LOC}\approx 0$ for $\phi<\phi_c^{\rm LOC}$ and $S_{\rm LOC}>0$ for $\phi>\phi_c^{\rm LOC}$ when the network is sufficiently large ($N \gg 1$). 

The random clustered network model introduced by Newman~\cite{newman2009random} generalizes the configuration model to incorporate clustering. Assume we are given the joint probability of $s$ and $t$, $p_{s,t}$, which represents the mean fraction of nodes with $s$ single edges and $t$ triangles in network realizations. Now, we start with $N$ nodes and, using $p_{s,t}$, we assign $s_i$ {\it edge stubs} and $t_i$ {\it triangle stubs} to each node $i$, under the constraint that $\sum_{i} s_i$ and $\sum_{i} t_i$ are multiples of 2 and 3, respectively. Given these stubs, we create a network by choosing pairs of edge stubs at random and joining them to make single edges, and choosing triples of triangle stubs at random and joining them to form triangles. This results in a random network where the number of single edges incident to each node and the number of triangles it participates in are distributed according to $p_{s,t}$, and where the nodes' degrees are essentially uncorrelated. Note that the total degree $k$ of a node with $s$ single edges and $t$ triangles is $k=s+2t$.

The clustering coefficient $C$ of this model is given by the generating functions~\cite{newman2009random}. First, we introduce the generating function $G_p(x,y)$ for the joint probability $p_{s,t}$:
\be
G_p(x,y)=\sum_{s=0}^\infty \sum_{t=0}^\infty p_{s,t} x^s y^t.
\ee
Since the full degree distribution $p_k$ is given by $p_k=\sum_{s,t} p_{s,t} \delta_{k, s+2t}$, where $\delta_{ij}$ is the Kronecker delta, the corresponding generating function is 
\be
G_{\rm tot}(z) = \sum_{k=0}^\infty p_k z^k = G_p(z,z^2).
\ee
The numbers $N_3$ of connected triplets and $N_\Delta$ of triangles are then given by the generating functions $G_p(x,y)$ and $G_{\rm tot}(z)$ \cite{newman2009random}:
\be
N_3 = N \sum_k \binom{k}{2} p_k = \frac{1}{2} N \frac{\partial^2 G_{\rm tot}(z)}{\partial z^2} \Big|_{z=1},
\ee
and
\be
3 N_\Delta = N \sum_{s,t} t p_{s,t} = N \frac{\partial G_p(x,y)}{\partial y} \Big|_{x=y=1}.
\ee
Since the clustering coefficient $C$ is defined as the ratio of 3 $\times$ the number of triangles to the number of connected triplets, we have the clustering coefficient of the random clustered network model as
\be
C = \frac{3 N_\Delta}{N_3} = 2 \frac{\partial G_p(x,y)}{\partial y} \Big|_{x=y=1} \Big/ \frac{\partial^2 G_{\rm tot}(z)}{\partial z^2} \Big|_{z=1}. \label{eq:clusteringCoefficient}
\ee

%%%%%%%%%%%%%%%%%%%%%%%%%%%%%%%%%%%%%%%%%%%%%%%%%%%%%%%%%%%%%%%%%%%%%%%%%%%
\section{Generating Function Analysis}
%%%%%%%%%%%%%%%%%%%%%%%%%%%%%%%%%%%%%%%%%%%%%%%%%%%%%%%%%%%%%%%%%%%%%%%%%%%

%%%%%%%%%%%%%%%%%%%%%%%%%
\subsection{Largest observable component}

First, we derive the LOC size for a random clustered network. In order to calculate the connected component statistics, we introduce two types of excess degree distributions \cite{newman2009random}: $q_{s,t}$, the probability that a node reached by traversing a single edge has $s+1$ single edges and $t$ triangles, and $r_{s,t}$, the probability that a node reached by traversing a triangle has $s$ single edges and $t+1$ triangles. For a random clustered network,
\be
q_{s,t}=\frac{s+1}{\langle s \rangle} p_{s+1,t}
\quad
{\rm and}
\quad
r_{s,t}=\frac{t+1}{\langle t \rangle} p_{s,t+1},
\ee
where $\langle s \rangle$ and $\langle t \rangle$ are the average values of $s$ and $t$, respectively. The generating functions for $q_{s,t}$ and $r_{s,t}$ are 
\be
G_q(x,y)=\sum_{s=0}^\infty \sum_{t=0}^\infty q_{s,t} x^s y^t=\frac{1}{\langle s \rangle} \frac{\partial G_p(x,y)}{\partial x}
\quad
{\rm and}
\quad
G_r(x,y)=\sum_{s=0}^\infty \sum_{t=0}^\infty r_{s,t} x^s y^t=\frac{1}{\langle t \rangle} \frac{\partial G_p(x,y)}{\partial y},
\ee
respectively.

In order to derive the LOC size, we need to obtain the following six quantities for adjacent nodes $i$ and $j$. 
\begin{enumerate}
\item
The probability $u_1$ that node $j$ is not a member of the LOC, given that node $i$ is D and is connected to node $j$ by a single edge. 
\item
The probability $u_2$ that node $j$ is not a member of the LOC, given that node $i$ is D and is connected to node $j$ by a triangle. This means $u_2^2$ is the probability that the two adjacent nodes forming a triangle with node $i$ are not members of the LOC, given that node $i$ is D.
\item
The probability $v_1$ that node $j$ is not a member of the LOC, given that node $i$ is I, node $j$ is not D (i.e., is I or U), and nodes $i$ and $j$ are connected by a single edge. 
\item
The probability $v_2$ that node $j$ is not a member of the LOC, given that node $i$ is I, node $j$ is not D, and nodes $i$ and $j$ are connected by a triangle. This means $v_2^2$ is the probability that the two adjacent nodes forming a triangle with node $i$ are not members of the LOC, given that node $i$ is I and neither neighbor is D.
\item
The probability $w_1$ that node $j$ is not a member of the LOC, given that node $i$ is I and is connected to node $j$ by a single edge.
\item
The probability $w_2^2$ that the two adjacent nodes forming a triangle with node $i$ are not members of the LOC, given that node $i$ is I. Note that the states of two adjacent nodes connected to a common I node are not independent~\footnote{This means the square root of $w_2^ 2$ does not correspond to any naturally-defined probability.}: one node cannot be U if the other node is D.
\end{enumerate}
Assuming we can approximate the network as a locally treelike graph incorporating triangles~\cite{newman2009random}, we can determine these quantities using the following self-consistent equations (see Appendix~\ref{sec:app} for details): 
\begin{subequations}
\begin{align}
u_1 &= \phi G_q(u_1, u_2^2) + \tphi G_q(w_1, w_2^2), \\
u_2^2 &= (\phi G_r(u_1, u_2^2) + \tphi G_r(w_1, w_2^2))^2, \\
v_1 &= G_q(\tphi, \tphi^2)+G_q(w_1, w_2^2)-G_q(\tphi v_1, \tphi^2 v_2^2), \\
v_2^2 &= (G_r(\tphi, \tphi^2)+G_r(w_1, w_2^2)-G_r(\tphi v_1, \tphi^2 v_2^2))^2, \\
w_1 &= \phi G_q(u_1,u_2^2) + \tphi v_1, \\
w_2^2 &= \phi^2 G_r(u_1,u_2^2)^2 +2 \phi \tphi G_r(u_1, u_2^2) G_r(w_1, w_2^2) + \tphi^2 v_2^2,
\end{align}
\label{eq:selfConsistentEqs}
\end{subequations}
where $\tphi=1-\phi$.

The fraction of nodes that are in the LOC, $S_{\rm LOC}$, is the probability that a randomly-chosen node belongs to the LOC. This is the sum of the probability that the node is D and belongs to the LOC, namely
\be
\phi (1- \sum_{s,t} p_{s,t} u_1^s u_2^{2t}), \nonumber
\ee
and the probability that it is I and belongs to the LOC, namely
\begin{eqnarray}
&& \tphi \sum_{s,t} p_{s,t} \Bigg[
\sum_{m=0}^s \binom{s}{m} \phi^m \tphi^{s-m} 
\sum_{n_1=0}^t \binom{t}{n_1} 
\sum_{n_2=0}^{n_1} \binom{n_1}{n_2} \phi^{2 n_2} (2 \phi \tphi)^{n_1-n_2} \tphi^{2(t-n_1)}
\nonumber \\
&& \times
\Big(1-G_q(u_1,u_2^2)^m v_1^{s-m} G_r(u_1,u_2^2)^{2n_2}(G_r(u_1,u_2^2)G_r(w_1,w_2^2))^{n_1-n_2}v_2^{2(t-n_1)} \Big)\Big(1-\delta_{m,0}\delta_{n_1,0}\Big) \Bigg]. \nonumber
\end{eqnarray}
This can be transformed to yield
\begin{eqnarray}
S_{\rm LOC} &=& \phi(1-G_p(u_1,u_2^2)) + \tphi(1-G_p(w_1, w_2^2)+G_p(\tphi v_1, \tphi^2 v_2^2)-G_p(\tphi,\tphi^2)) \nonumber \\
&=& 1-\phi G_p(u_1,u_2^2) - \tphi G_p(w_1,w_2^2) + \tphi G_p(\tphi v_1, \tphi v_2^2) - \tphi G_p(\tphi, \tphi^2). \label{eq:sLOC}
\end{eqnarray}
Equations (\ref{eq:selfConsistentEqs}) and (\ref{eq:sLOC}) enable us to evaluate $S_{\rm LOC}$ and also obtain the critical probability $\phi_c^{\rm LOC}$ numerically.

%%%%%%%%%%%%%%%%%%%%%%%%%
\begin{figure}[t]
\begin{center}
\includegraphics[width=.50\textwidth]{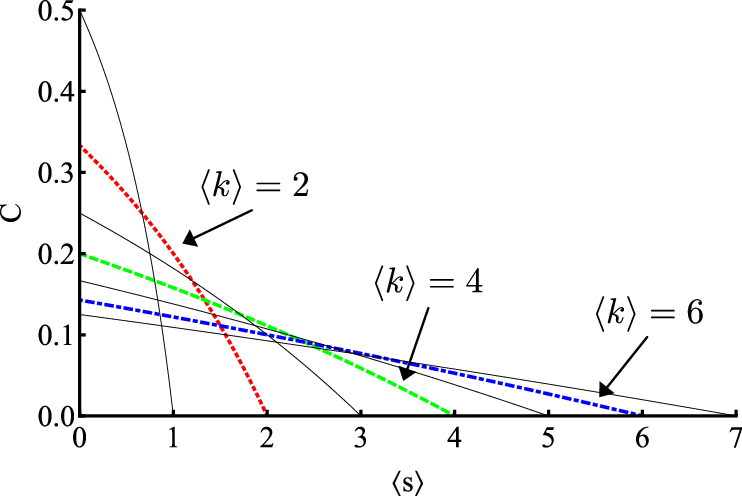}
\caption{
Clustering coefficient $C$ as a function of $\langle s \rangle$ ($0 \le \langle s \rangle \le \langle k \rangle$) for random clustered networks with $\langle k \rangle=1,2,\cdots,7$. The dashed red, green, and blue lines represent $\langle k \rangle=2$, 4, and 6, respectively. Here, all lines are drawn using Eqs.~(\ref{eq:clusteringCoefficient}) and (\ref{eq:generaitingFunctionDoublePoisson}).
}
\label{fig:clusteringCoefficient}
\end{center}
\end{figure}
%%%%%%%%%%%%%%%%%%%%%%%%%

Now, we work through a simple example in order to inspect the effect of clustering on network observability. Consider a random clustered network with the doubly-Poisson distribution 
\be
p_{s,t} = e^{-\langle s \rangle} \frac{\langle s \rangle^s}{s!} e^{-\langle t \rangle} \frac{\langle t \rangle^t}{t!}, \label{eq:doublePoisson}
\ee
which yields $\langle k \rangle=\langle s \rangle+2\langle t \rangle$. In this case, the generating functions can be simplified to
\be
G_p(x,y)=G_q(x,y)=G_r(x,y)=e^{\langle s \rangle (x-1)}e^{\langle t \rangle (y-1)}, \label{eq:generaitingFunctionDoublePoisson}
\ee
and thus the clustering coefficient $C$ is given by $C=2 \langle t \rangle/(2\langle t \rangle+(\langle s \rangle+2\langle t \rangle)^2)$. Figure~\ref{fig:clusteringCoefficient} plots the clustering coefficient $C$ as a function of $\langle s \rangle$ for several values of $\langle k \rangle$, showing that it decreases as $\langle s \rangle$ increases for fixed $\langle k \rangle$. It also shows that larger values of $\langle k \rangle$ give lower maximum clustering coefficient values.

%%%%%%%%%%%%%%%%%%%%%%%%%
\begin{figure}[t]
\begin{center}
\includegraphics[width=.50\textwidth]{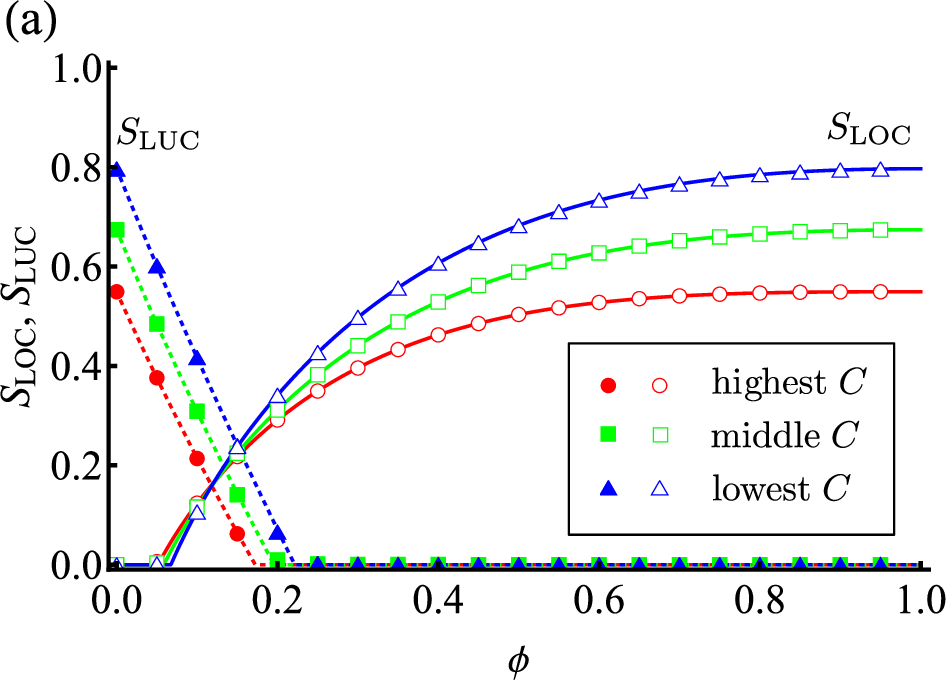}
\includegraphics[width=.50\textwidth]{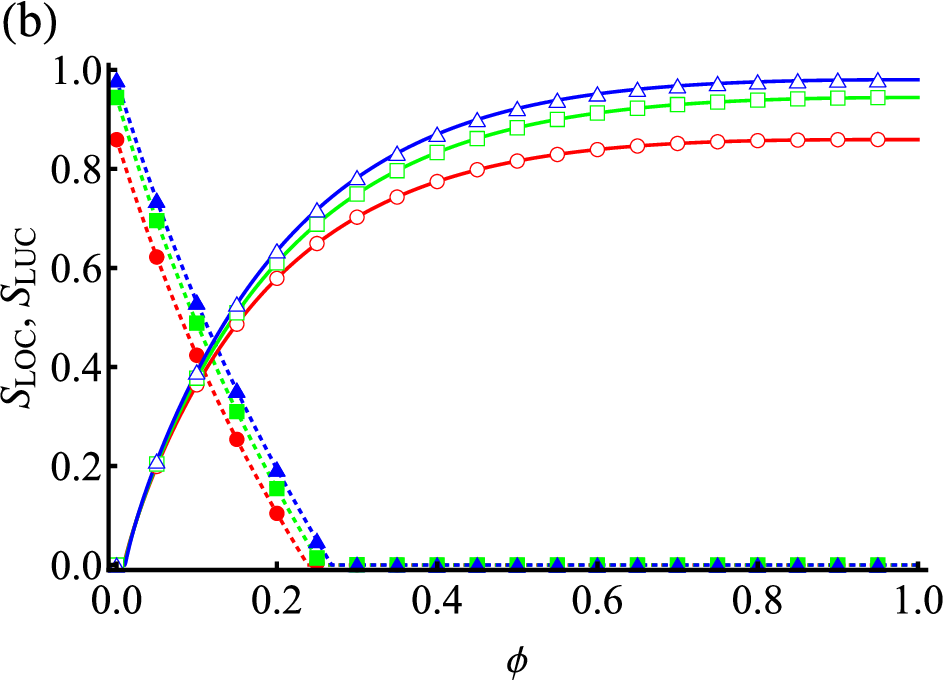}
\includegraphics[width=.50\textwidth]{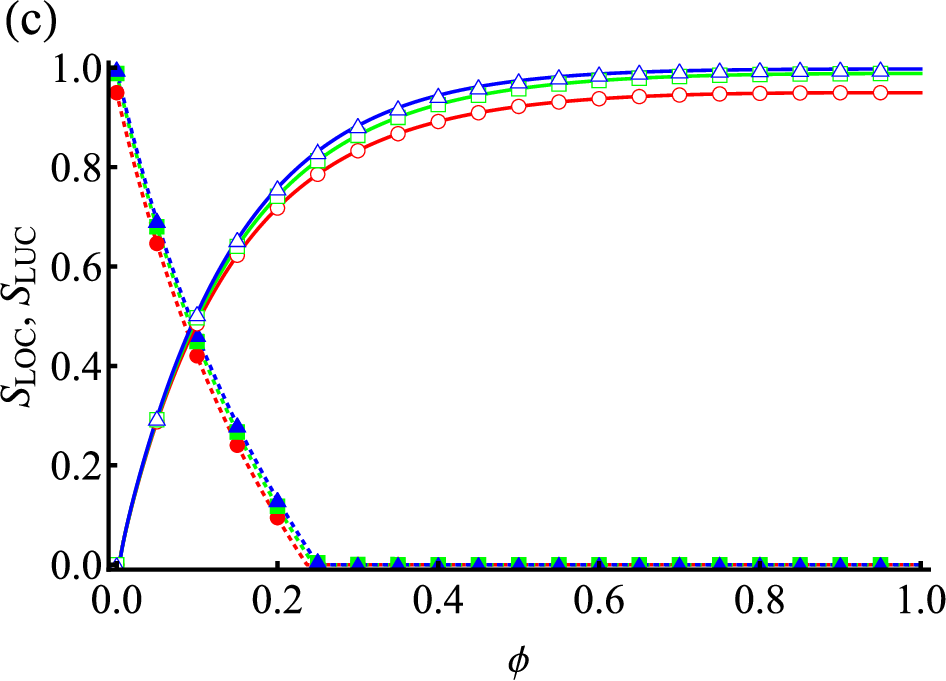}
\caption{
Normalized LOC and LUC sizes, $S_{\rm LOC}$ and $S_{\rm LUC}$, for random clustered networks with doubly-Poisson distributions when (a) $\langle k \rangle=2$, (b) $\langle k \rangle=4$, and (c) $\langle k \rangle=6$. 
The red circles, green squares, and blue triangles show the results of Monte Carlo simulations using $N=10^5$ nodes, for the highest possible $C$ ($\langle s \rangle=0$), middle $C$ ($\langle s \rangle=\langle k \rangle/2$), and lowest possible $C$ ($\langle s \rangle=\langle k \rangle$), respectively. 
The open and filled symbols indicate $S_{\rm LOC}$ and $S_{\rm LUC}$, respectively, while the solid and dotted lines represent the corresponding analytical results. 
These values were calculated by constructing $10^2$ network realizations, carrying out a single run on each, and averaging the results. 
For the $\langle k \rangle=2$ case, see also Fig.~\ref{fig:narrow} in Appendix~\ref{sec:app2}, which plots $S_{\rm LOC}$ in the small-$\phi$ region.
}
\label{fig:orderParameter}
\end{center}
\end{figure}
%%%%%%%%%%%%%%%%%%%%%%%%%

Figures~\ref{fig:orderParameter} (a)--(c) plot the normalized LOC sizes, $S_{\rm LOC}$, of random clustered networks with the highest possible $C$ ($\langle s \rangle=0$), middle $C$ ($\langle s \rangle=\langle k \rangle/2$), and lowest possible $C$ ($\langle s \rangle=\langle k \rangle$), for several values of $\langle k \rangle$. They compare the analytical results (solid lines) with the Monte Carlo results (open symbols). For each $\phi$ value, we carried out Monte Carlo simulations of a single random sensor placement run on each of $10^2$ network realizations consisting of $10^5$ nodes, and calculated the average normalized LOC size. As the figures show, the analytical and simulation results are in good agreement in all cases.

Figure~\ref{fig:orderParameter} (a) shows the results for $\langle k \rangle=2$. Both the analytical and numerical results show that the LOC depends on the value of $C$. 
In particular, higher clustering makes it easier for observable nodes to connect to each other through more redundant paths and therefore the critical probability $\phi_c^{\rm LOC}$ decreases, even though the giant component size (which is equivalent to the LOC size at $\phi=1$) is smaller as the network is more clustered.
For the present case, it seems natural to presume $\phi_{c,{\rm triangles}}^{\rm LOC}<\phi_c^{\rm LOC}<\phi_{c,{\rm edges}}^{\rm LOC}$ for $0<\langle s \rangle<\langle k \rangle$, where $\phi_{c,{\rm triangles}}^{\rm LOC}$ and $\phi_{c,{\rm edges}}^{\rm LOC}$ are the critical LOC probabilities for $\langle s \rangle=0$ (i.e., no single edges) and $\langle s \rangle=\langle k \rangle$ (i.e., no triangles), respectively.

Figure~\ref{fig:orderParameter} (a) actually tells us that a network's clustering affects its observability, although the effect is quantitatively small. Figures~\ref{fig:orderParameter} (b) and (c) show similar plots for larger $\langle k \rangle$ values. 
These indicate that the critical probability $\phi_c^{\rm LOC}$ is very close to zero and $S_{\rm LOC}$ quickly increases as $\phi$ increases from zero, irrespective of $C$.
That is, the degree of clustering in a network has no obvious effect on the observability transition when $\langle k \rangle$ is large.

%%%%%%%%%%%%%%%%%%%%%%%%%
\subsection{Largest unobservable component}

There is also a phase transition associated with the LUC~\cite{allard2014coexistence}. The effect of clustering on network observability may be reflected in the LUC size rather than the LOC size. Here, we derive the LUC size using the generating functions.

First, we denote the probability that the node reached by traversing a single edge connecting to a non-D node (i.e., an I or U node) is U by $\psi_{q}$, and the probability that the node reached by traversing a triangle from a non-D node is U by $\psi_{r}$. These probabilities can be naturally given as
\be
\psi_{q} = \tphi G_q(\tphi, \tphi^2) 
\quad {\rm and} \quad
\psi_{r} = \tphi G_r(\tphi, \tphi^2).
\ee
Now, we consider the joint probability $P(m,n|{\rm U})$ that a node in a U component (a connected component of U nodes) has $m$ neighbors connected by single edges and $n$ neighbors in triangles within that component. We begin with the probability $P({\rm U},m,n|s,t)$ that a randomly-chosen node belongs to a U component and has $m$ U neighbors connected by single edges and $n$ U neighbors in triangles, given that it has $s$ single edges and $t$ triangles in the original clustered network. Noting that all neighbors of a U node must be I or U, we have
\be
P({\rm U},m,n|s,t)=\tphi \binom{s}{m} (\psi_{q})^m (\tphi-\psi_{q})^{s-m} \binom{2t}{n} (\psi_{r})^n (\tphi-\psi_{r})^{2t-n}.
\ee
The probability $P({\rm U},m,n)$ that a randomly-chosen node belongs to a U component and has $m$ U neighbors connected by single edges and $n$ U neighbors in triangles is
\be
P({\rm U},m,n)=\sum_{s=m}^\infty \sum_{2t=n}^\infty p_{s,t} P({\rm U},m,n|s,t),
\ee
and the probability $P({\rm U})$ that a randomly-chosen node belongs to a U component is 
\be
P({\rm U})=\sum_{m=0}^\infty \sum_{n=0}^\infty P({\rm U},m,n)=\tphi G_p(\tphi, \tphi^2).
\ee
Since $P(m,n|{\rm U})=P({\rm U},m,n)/P({\rm U})$, we can represent the conditional joint probability $P(m,n|{\rm U})$ as 
\begin{eqnarray}
P(m,n|{\rm U})
&=&\frac{1}{P({\rm U})} \sum_{s=m}^\infty \sum_{2t=n}^\infty p_{s,t} P({\rm U},m,n|s,t) \nonumber \\
&=&\frac{1}{P({\rm U})} \sum_{s=m}^\infty \sum_{2t=n}^\infty p_{s,t} \tphi \binom{s}{m} (\psi_{q})^m (\tphi-\psi_{q})^{s-m} \binom{2t}{n} (\psi_{r})^n (\tphi-\psi_{r})^{2t-n}.
\end{eqnarray}

Now, we can introduce the generating functions for the probability distributions of U components. First, the generating function $F_p(x,y)$ for the joint probability $P(m,n|{\rm U})$ is 
\begin{eqnarray}
F_p(x,y) 
&=& \sum_{m=0}^\infty \sum_{n=0}^\infty P(m,n|{\rm U}) x^m y^n \nonumber \\
&=&\frac{\tphi}{P({\rm U})} \sum_{m=0}^\infty \sum_{n=0}^\infty p_{s,t} (\psi_{q} x+\tphi-\psi_{q})^s (\psi_{r} x+\tphi-\psi_{r})^{2t}
\nonumber \\
&=&\frac{1}{G_p(\tphi,\tphi^2)} G_p(\psi_{q} x+\tphi-\psi_{q}, (\psi_{r} y+\tphi-\psi_{r})^2).
\end{eqnarray}

To derive the LUC size, we also introduce two additional joint distributions of U components: $P_q(m,n|{\rm U})$, the probability that a node, belonging to a U component and reached by traversing a single edge, has $m+1$ neighbors connected by single edges and $n$ neighbors in triangles within that component, and $P_r(m,n|{\rm U})$, the probability that a node, belonging to a U component and reached by traversing a triangle, has $m$ neighbors connected by single edges and $n+1$ neighbors in triangles within in that component. Similarly to $F_p(x,y)$, we can obtain the generating functions for the joint distributions $P_q(m,n|{\rm U})$ and $P_r(m,n|{\rm U})$ as
\be
F_q(x,y) =\frac{1}{G_q(\tphi,\tphi^2)} G_q(\psi_{q} x+\tphi-\psi_{q}, (\psi_{r} y+\tphi-\psi_{r})^2)
\ee
and
\be
F_r(x,y) =\frac{1}{G_r(\tphi,\tphi^2)} G_r(\psi_{q} x+\tphi-\psi_{q}, (\psi_{r} y+\tphi-\psi_{r})^2),
\ee
respectively.

Next, we turn to percolation analysis of U components. If we denote the probability that a U node reached by traversing a single edge is not a member of the LUC by $u_{\rm U}$ and the corresponding probability for a U node reached by traversing a triangle by $v_{\rm U}$, then we have the following self-consistent equations for $u_{\rm U}$ and $v_{\rm U}$:
\be
u_{\rm U} = F_q(u_{\rm U}, v_{\rm U}) 
\quad {\rm and} \quad
v_{\rm U} = F_r(u_{\rm U}, v_{\rm U}).
\ee
Since the normalized LUC size $S_{\rm LUC}$ is one minus the probability that a randomly-chosen node is U but not connected to the LUC, we have
\begin{eqnarray}
S_{\rm LUC} 
&=& P({ \rm U}) \Big( 1- \sum_{m,n} P(m,n | {\rm U})u_{\rm U}^m v_{\rm U}^{n} \Big) \nonumber \\
&=& \tphi G_p(\tphi, \tphi^2) (1-F_p(u_{\rm U}, v_{\rm U})).
\label{eq:sLUC}
\end{eqnarray}
These equations allow us to calculate the normalized LUC size $S_{\rm LUC}$. We can then obtain the critical probability $\phi_c^{\rm LUC}$ numerically, such that $S_{\rm LUC}>0$ for $\phi<\phi_c^{\rm LUC}$ and $S_{\rm LUC}=0$ for $\phi>\phi_c^{\rm LUC}$ in the limit $N \to \infty$.

Figures~\ref{fig:orderParameter} (a)--(c) show the normalized LUC sizes for random clustered networks with doubly-Poisson distributions (\ref{eq:doublePoisson}). As with the LOC, we find that the analytical results match the simulation results precisely. We also find that $S_{\rm LUC}$ depends on the clustering coefficient $C$ in all cases shown. Higher clustering reduces $S_{\rm LUC}$ (possibly due to suppression of the giant component), indicating that the critical probability $\phi_c^{\rm LUC}$ also decreases with increasing $C$. Thus, we again expect for the present case that $\phi_{c,{\rm triangles}}^{\rm LUC}<\phi_c^{\rm LUC}<\phi_{c,{\rm edges}}^{\rm LUC}$ for $0 < \langle s \rangle < \langle k \rangle$, where $\phi_{c,{\rm triangles}}^{\rm LUC}$ and $\phi_{c,{\rm edges}}^{\rm LUC}$ are the critical LUC probabilities for $\langle s \rangle=0$ and $\langle s \rangle=\langle k \rangle$, respectively.

%%%%%%%%%%%%%%%%%%%%%%%%%
\begin{figure}[t]
\begin{center}
\includegraphics[width=.50\textwidth]{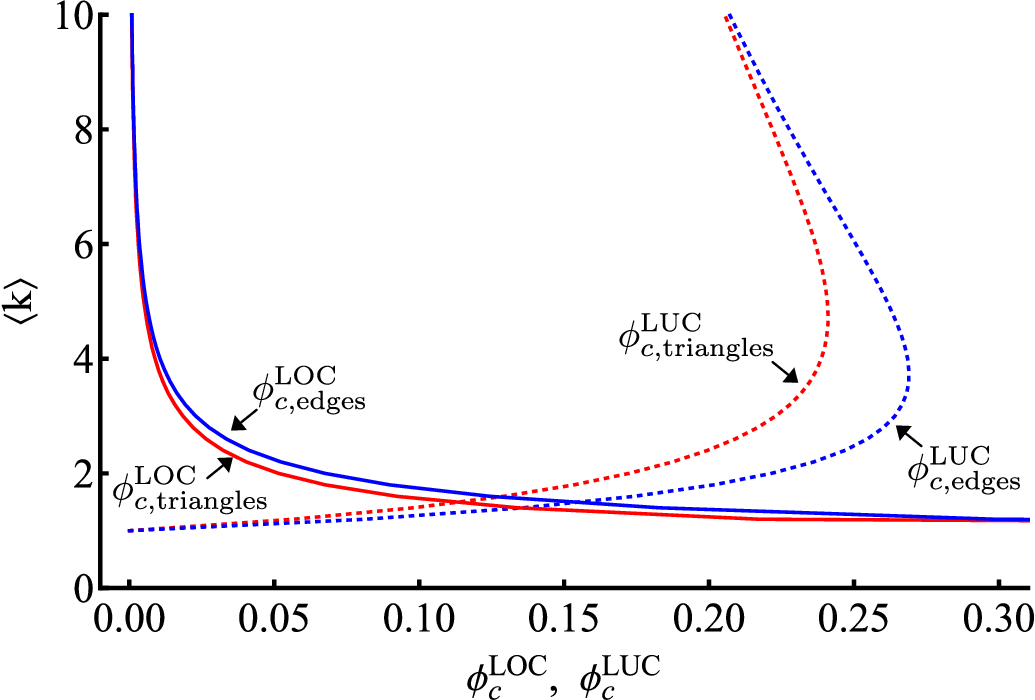}
\caption{
Bounds on the critical probabilities, $\phi_c^{\rm LOC}$ and $\phi_c^{\rm LUC}$, for random clustered networks with doubly-Poisson distributions. The solid and dotted red lines represent the critical probabilities for $\langle s \rangle=0$ and $\langle t \rangle=\langle k \rangle/2$, i.e., $\phi_{c,{\rm triangles}}^{\rm LOC}$ and $\phi_{c,{\rm triangles}}^{\rm LUC}$, respectively. Likewise, the solid and dotted blue lines represent the critical probabilities for $\langle s \rangle=\langle k \rangle$ and $\langle t \rangle=0$, i.e. $\phi_{c,{\rm edges}}^{\rm LOC}$ and $\phi_{c,{\rm edges}}^{\rm LUC}$, respectively. For a fixed value of $\langle k \rangle=\langle s \rangle+2\langle t \rangle$, the critical probabilities for other combinations of $\langle s \rangle>0$ and $\langle t \rangle>0$ lie between the red and blue lines.
}
\label{fig:phaseDiagram}
\end{center}
\end{figure}
%%%%%%%%%%%%%%%%%%%%%%%%%

%%%%%%%%%%%%%%%%%%%%%%%%%
\subsection{Bounds on $\phi_c^{\rm LOC}$ and $\phi_c^{\rm LUC}$}

It is difficult to obtain expressions for the critical probabilities $\phi_c^{\rm LOC}$ and $\phi_c^{\rm LUC}$ for random clustered networks, because the self-consistent equations that determine $S_{\rm LOC}$ and $S_{\rm LUC}$ are complicated. Instead, we discuss bounds on them, supposing $\phi_{c,{\rm triangles}}^{\rm LOC}<\phi_c^{\rm LOC}<\phi_{c,{\rm edges}}^{\rm LOC}$ and $\phi_{c,{\rm triangles}}^{\rm LUC}<\phi_c^{\rm LUC}<\phi_{c,{\rm edges}}^{\rm LUC}$ for $0 < \langle s \rangle < \langle k \rangle$. Figure~\ref{fig:phaseDiagram} shows the critical probability bounds for random clustered networks with doubly-Poisson distributions. Here, we find that clustering's effect on $\phi_c^{\rm LOC}$ gradually disappears as the average degree $\langle k \rangle$ increases, considering the difference between $\phi_{c,{\rm edges}}^{\rm LOC}$ and $\phi_{c,{\rm triangles}}^{\rm LOC}$. 
This is also true for the LUC.
These results lead to the conclusion for random clustered networks that clustering has an almost negligible effect on the network observability transition when the average degree is large ($\langle k \rangle >10$).

%%%%%%%%%%%%%%%%%%%%%%%%%
\begin{figure}[t]
\begin{center}
\includegraphics[width=.50\textwidth]{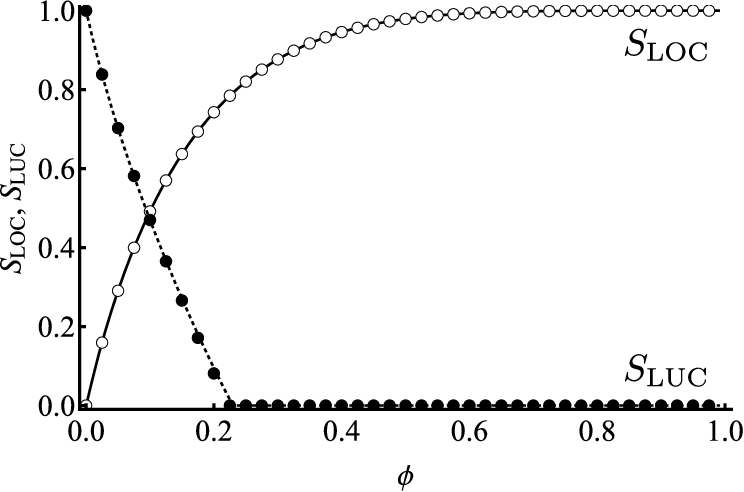}
\caption{
Normalized LOC and LUC sizes, $S_{\rm LOC}$ and $S_{\rm LUC}$, for a random clustered network with a power-law triangle distribution $p_{s,t}=t^{-\gamma}/\sum_{t'=t_{\rm min}}^{t_{\rm max}} {t'}^{-\gamma}$, where $\gamma=3.0$, $t_{\rm min}=2$, $t_{\rm max}=10^2$, and $p_{s,t}$ is independent of $s$ (meaning there are no single edges), and for a degree-preserving randomized network. The former clustering coefficient is $C \approx 0.082$. The open and filled symbols represent $S_{\rm LOC}$ and $S_{\rm LUC}$, respectively, for the random clustered scale-free network, while the solid and dotted lines represent the corresponding randomized network values, all calculated from Eqs.~(\ref{eq:sLOC}) and (\ref{eq:sLUC}). Monte Carlo results for these networks are also in good agreement with each other (not shown).
}
\label{fig:Scale-freeNetwork}
\end{center}
\end{figure}
%%%%%%%%%%%%%%%%%%%%%%%%%

%%%%%%%%%%%%%%%%%%%%%%%%%%%%%%%%%%%%%%%%%%%%%%%%%%%%%%%%%%%%%%%%%%%%%%%%%%%
\section{Summary and Discussion}
%%%%%%%%%%%%%%%%%%%%%%%%%%%%%%%%%%%%%%%%%%%%%%%%%%%%%%%%%%%%%%%%%%%%%%%%%%%

In this paper, we have investigated the relationship between a network's clustering and its observability. We have derived the sizes of the largest observable component (LOC) and largest unobservable component (LUC) in the random clustered network model. We have demonstrated, both analytically and numerically, that the clustered network's structure does affect its observability transitions. More highly-clustered structures make it easier to form macroscopic LOCs, so the associated critical probability decreases as the clustering coefficient $C$ increases. The clustering structure also affects the LUC, with larger $C$ values reducing the critical probability in this case as well. Our theoretical results indicate that although the network's clustering influences its observability transition, the effect is weak. We also gave numerical bounds on the critical probabilities, $\phi_c^{\rm LOC}$ and $\phi_c^{\rm LUC}$, for random clustered networks, showing that the effect of a network's clustering becomes almost negligible unless its average degree is small.

It should be mentioned that this weak or negligible effect of a network's clustering on its observability holds true for scale-free networks as well. Comparing the normalized LOC and LUC sizes for a random clustered scale-free network and a degree-preserving randomized network, we found that the results matched exactly, implying that clustering has a negligible effect on observability in scale-free networks (Fig.~\ref{fig:Scale-freeNetwork}). This is consistent with a previous study by Yang and Radicchi~\cite{yang2016observability}, who found that the observability of real-world networks is well-described by a message-passing approach that assumes a locally-treelike approximation, even when the network has a large clustering coefficient.

In this study, we used a random clustered network model introduced by~\cite{newman2009random,miller2009percolation}, which has the limitation that possible values of $C$ become small as the average degree increases.
Other network models with further tunable clustering coefficients have been proposed~\cite{holme2002growing,klemm2002highly,volz2004random,serrano2005tuning,trapman2007analytical,mann2021random}.
It may be interesting to investigate whether this study's findings hold for those networks as well, although we expect they too will exhibit weak or negligible dependence of clustering on observability.

There are related topics to the network observability model. For example, the problem of finding the smallest set of directly-observable nodes that make the entire network observable is known as the minimum dominating set problem~\cite{molnar2014dominating,molnar2015building,zhao2015statistical}. Other models, similar to the network observability model, may be relevant to behavior on social networks, e.g., vaccinations by observers~\cite{takaguchi2014suppressing} or quarantine measures on the spread of epidemics~\cite{hasegawa2017efficiency}. It may be interesting to clarify how network clustering affects such social models, based on the findings of this study.

%%%%%%%%%%%%%%%%%%%%%%%%%%%%%%%%%%%%%%%%%%%%%%%%%%%%%%%%%%%%%%%%%%%%%%%%%%%
\section*{Acknowledgements}
The authors thank Shogo Mizutaka for helpful discussions. 
This work was supported by JSPS KAKENHI Grant Numbers JP15K17716, JP16H03939, and JP18KT0059.
%%%%%%%%%%%%%%%%%%%%%%%%%%%%%%%%%%%%%%%%%%%%%%%%%%%%%%%%%%%%%%%%%%%%%%%%%%%

%\bibliography{observability}

\appendix

%%%%%%%%%%%%%%%%%%%%%%%%%
\begin{figure}[t]
\begin{center}
\includegraphics[width=0.90\textwidth]{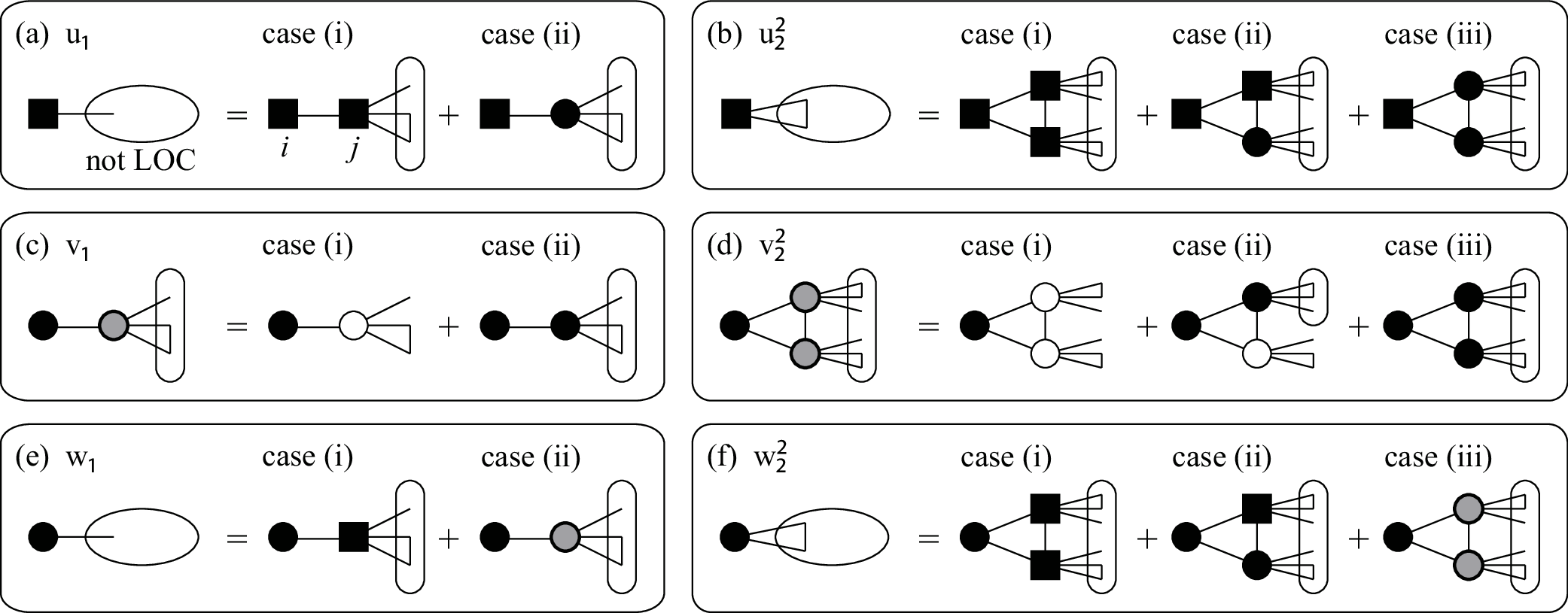}
\caption{
Illustration of the contributions to the self-consistent equations for (a) $u_1$, (b) $u_2^2$, (c) $v_1$, (d) $v_2^2$, (e) $w_1$, and (f) $w_2^2$. The black squares, black circles, and white circles represent directly observable (D), indirectly observable (I), and unobservable (U) nodes, respectively, while the gray circles represent nodes that are not D, i.e., are I or U.
}
\label{fig:schematic}
\end{center}
\end{figure}
%%%%%%%%%%%%%%%%%%%%%%%%%

%%%%%%%%%%%%%%%%%%%%%%%%%%%%%%%%%%%%%%%%%%%%%%%%%%%%%%%%%%%%%%%%%%%%%%%%%%%
\section{Derivation of the self-consistent equations \label{sec:app}}
%%%%%%%%%%%%%%%%%%%%%%%%%%%%%%%%%%%%%%%%%%%%%%%%%%%%%%%%%%%%%%%%%%%%%%%%%%%

In this appendix, we derive the self-consistent equations for $u_1$, $u_2^2$, $v_1$, $v_2^2$, $w_1$, and $w_2^2$. For convenience, we call a node adjacent to a randomly-chosen node $i$ an edge-neighbor when it is connected to node $i$ by a single edge, and a triangle-neighbor when it is connected to node $i$ as part of a triangle.

First, the probability $u_1$ that an edge-neighbor $j$ is not a member of the LOC, given that node $i$ is D, involves two cases (Fig.~\ref{fig:schematic} (a)). Because node $j$ is connected to a D node $i$, it is D with probability $\phi$ and I otherwise. In case (i), the edge-neighbor $j$ is D but is not connected to the LOC via its excess neighbors (i.e., neighbors other than node $i$). An edge-neighbor has $s$ edge-neighbors and $t$ triangles ($2t$ triangle-neighbors) with probability $q_{s,t}$. By definition, an edge-neighbor and a triangle connected to a D node are not members of the LOC with probabilities $u_1$ and $u_2^2$, respectively. Thus, this case contributes $\phi \sum_{s,t} q_{s,t} u_1^s u_2^{2t} = \phi G_q(u_1, u_2^2)$. In case (ii), node $j$ is I, which happens with probability $\tphi$, but is not connected to the LOC via its excess neighbors. Noting that an edge-neighbor and a triangle connected to an I node $j$ are not members of the LOC with probabilities $w_1$ and $w_2^2$, respectively, the contribution of this case is $\tphi \sum_{s,t} q_{s,t} w_1^s w_2^{2t} =\tphi G_q(w_1, w_2^2)$. Putting this together, we obtain the self-consistent equation for $u_1$:
\be
u_1 = \phi G_q(u_1, u_2^2) + \tphi G_q(w_1, w_2^2).
\ee

The probability $u_2^2$ that two triangle-neighbors forming a triangle with node $i$ are not members of the LOC, given that node $i$ is D, involves three cases (Fig.~\ref{fig:schematic} (b)): (i) both triangle-neighbors are D, (ii) one is D and the other is I, and (iii) both are I. Noting that for a triangle-neighbor the joint distribution of excess edge-neighbors and triangles is given by $r_{s,t}$, these contributions are $\phi^2 G_r(u_1, u_2^2)^2$, $\phi \tphi G_r(u_1, u_2^2)G_r(w_1, w_2^2)$, and $\tphi^2 G_r(w_1, w_2^2)^2$, respectively. Combining these, the self-consistent equation for $u_2^2$ is
\be
u_2^2 = (\phi G_r(u_1, u_2^2) + \tphi G_r(w_1, w_2^2))^2.
\ee

Next, the probability $v_1$ that an edge-neighbor $j$ is not a member of the LOC, given that node $i$ is I and $j$ is not D, involves two cases (Fig.~\ref{fig:schematic} (c)). In case (i), node $j$ is U; this occurs with probability $G_q(\phi, \tphi)$. In case (ii), node $j$ is I but is not a member of the LOC via its excess neighbors. Noting that node $j$ can be I if at least one excess neighbor is D, given that node $i$ is I, this probability is $G_q(w_1, w_2^2)-G_q(\tphi v_1, \tphi^2 v_2^2)$, where $G_q(w_1, w_2^2)$ is the probability that I node $j$ is not connected to the LOC and $G_q(\tphi v_1, \tphi^2 v_2^2)$ is the probability that it is not a member of the LOC and none of its neighbors are D. Combining these contributions, the self-consistent equation for $v_1$ is
\be
v_1 = G_q(\tphi, \tphi^2)+G_q(w_1, w_2^2)-G_q(\tphi v_1, \tphi^2 v_2^2).
\ee

The probability $v_2^2$ that two triangle-neighbors forming a triangle with node $i$ are not members of the LOC, given that node $i$ is I and neither triangle-neighbor is D, involves three cases (Fig.~\ref{fig:schematic} (d)): (i) both triangle-neighbors are U, (ii) one is I and the other is U, and (iii) both are I. Noting that the states of these triangle-neighbors are independent, these contributions are $G_r(\tphi, \tphi^2)^2$, $2G_r(\tphi, \tphi^2)(G_r(w_1, w_2^2)-G_r(\tphi v_1, \tphi^2 v_2^2))$, and $(G_r(w_1, w_2^2)-G_r(\tphi v_1, \tphi^2 v_2^2))^2$, respectively. Thus, the self-consistent equation for $v_2^2$ is
\be
v_2^2 = (G_r(\tphi, \tphi^2)+G_r(w_1, w_2^2)-G_r(\tphi v_1, \tphi^2 v_2^2))^2.
\ee

The probability $w_1$ that an edge-neighbor $j$ is not a member of the LOC, given that node $i$ is I, involves two cases (Fig.~\ref{fig:schematic} (e)), depending on whether or not node $j$ is D. The contribution of the first case (node $j$ is D) is $\phi G_q(u_1,u_2^2)$, while, for the latter case, node $j$ is not D with probability $\tphi$. The probability that node $j$ is not a member of the LOC, given that node $i$ is I and $j$ is not D, is $v_1$ by definition, so the self-consistent equation for $w_1$ is
\be
w_1 = \phi G_q(u_1,u_2^2) + \tphi v_1.
\ee

Finally, the probability $w_2^2$ that two triangle-neighbors of node $i$ are not members of the LOC, given that node $i$ is I, involves three cases (Fig.~\ref{fig:schematic} (f)): (i) both triangle-neighbors are D, (ii) one is D and the other is I, and (iii) neither is D. Here, we note that the states of the two triangle-neighbors are not independent: if one is D, the other cannot be U, and thus is D with probability $\phi$ or I with probability $\tphi$. Since the contributions of cases (i), (ii), and (iii) are $\phi^2 G_r(u_1,u_2^2)^2$, $2 \phi \tphi G_r(u_1, u_2^2) G_r(w_1, w_2^2)$, and $\tphi^2 v_2^2$, respectively, we have that the self-consistent equation for $w_2^2$ is
\be
w_2^2 = \phi^2 G_r(u_1,u_2^2)^2 +2 \phi \tphi G_r(u_1, u_2^2) G_r(w_1, w_2^2) + \tphi^2 v_2^2.
\ee

%%%%%%%%%%%%%%%%%%%%%%%%%
\begin{figure}[t]
\begin{center}
\includegraphics[width=.50\textwidth]{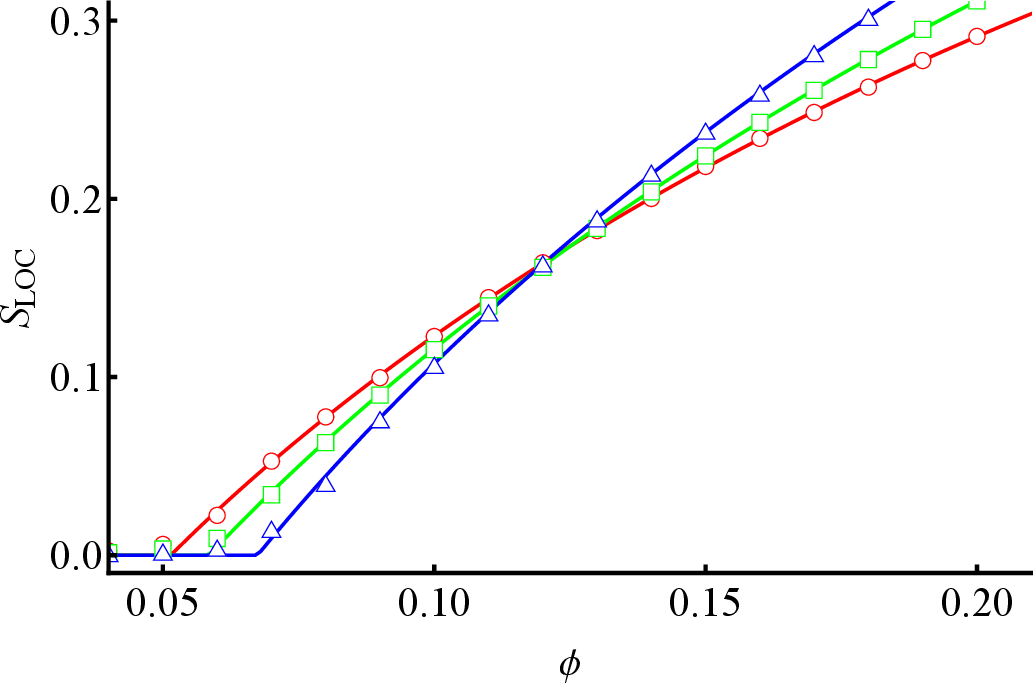}
\caption{
Normalized LOC size $S_{\rm LOC}$ for a random clustered network with a doubly-Poisson distribution and $\langle k \rangle=2$. The red circles, green squares, and blue triangles show the results of Monte Carlo simulations for the highest possible $C$ ($\langle s \rangle=0$), intermediate $C$ ($\langle s \rangle=\langle k \rangle/2$), and lowest possible $C$ ($\langle s \rangle=\langle k \rangle$), respectively, while the solid lines show the corresponding analytical results, calculated using Eq.~(\ref{eq:sLOC}). The simulation results are average values over $10^2$ network realizations consisting of $N=10^5$ nodes, and a single sensor placement run was carried out for each one.
}
\label{fig:narrow}
\end{center}
\end{figure}
%%%%%%%%%%%%%%%%%%%%%%%%%

\section{Plot of $S_{\rm LOC}$ in the small-$\phi$ region \label{sec:app2}}

Following Fig.~\ref{fig:orderParameter} (a), we compare the analytical estimates and Monte Carlo simulation results for $S_{\rm LOC}$ in the small-$\phi$ region, in order to demonstrate more clearly that stronger clustering leads to lower critical probabilities. Figure~\ref{fig:narrow} shows analytical (lines) and numerical (symbols) $S_{\rm LOC}$ values for a random clustered network with a doubly-Poisson distribution and $\langle k \rangle=2$. Here, we can see that the analytical and numerical results match perfectly, confirming that $\phi_c^{\rm LOC}$ does indeed decrease as $C$ increases.

\end{document}